\documentstyle[aps,prl,epsfig,floats,preprint,tighten]{revtex}

\begin{document}

\hyphenation{wave-guide}
\draft

\date{\today}
\title{Dynamic effect of phase conjugation on wave localization}
\author{K. J. H. van Bemmel, M. Titov, and C. W. J. Beenakker}
\address{Instituut-Lorentz, Universiteit Leiden, P.O. Box 9506, 2300 RA
Leiden, The Netherlands}

\widetext

\maketitle

\begin{abstract}
We investigate what would happen to the time dependence of a pulse reflected by 
a disordered single-mode waveguide, if it is closed at one end not by an ordinary mirror but 
by a phase-conjugating mirror. We find that the waveguide acts like
a virtual cavity with resonance frequency equal to the working frequency $\omega_0$
of the phase-conjugating mirror. The decay in time of the average power 
spectrum of the reflected pulse is delayed for
frequencies near $\omega_0$. In the presence of localization 
the resonance width 
is $\tau_s ^{-1} \exp(-L/l)$, with $L$ the length of the waveguide, $l$ the
mean free path, and $\tau_s$ the scattering time. Inside this frequency range
the decay of the average power spectrum is delayed up to times $t\simeq \tau_s \exp(L/l)$.
\end{abstract}

\vspace{0.5cm}


%\chapter{Dynamic effect of phase conjugation on wave localization}
%\chaptermark{Dynamic effect of phase conjugation \ldots}
%\label{pcm}

\section{Introduction}

The reflection of a wave pulse by a random medium provides insight into the dynamics of localization \cite{She90,Papan,Fouque,revBeen}. The reflected amplitude contains rapid fluctuations over a broad range of frequencies, with a slowly
decaying envelope. The power spectrum $a(\omega ,t)$ characterizes the decay in time $t$ of the envelope at frequency 
$\omega$. In an infinitely long waveguide (with $N$ propagating modes), the signature of localization \cite{Whi87,Titov},

\begin{equation} \label{introform}
\mbox{} \Big\langle a(\omega,t) \Big\rangle \propto t^{-2}  \hspace{0.5cm} {\mathrm{for}} \hspace{0.3cm} t \gg N^2\tau_s,
\end{equation}
is a quadratic decay of the disorder-averaged power spectrum $\langle a \rangle$, 
that sets in after $N^2$ scattering times $\tau_s$. 

The decay ($\ref{introform}$) still holds over a broad range of times if the length $L$ of the waveguide is finite, but much greater than the localization length $\xi=(N+1)l$ (with $l=c\tau_s$ the mean free path).
What changes is that for exponentially large times $t \gg \tau_s \exp(L/l)$ the quadratic decay becomes more rapid
$\propto \exp(-{\mathrm{constant}} \times \ln ^2 t)$. This is the celebrated log-normal tail \cite{dynAltshuler,AKL,Muzy,Bolton,Mirlin}.
We may assume that the finite length of the waveguide is realized by terminating one end by a perfectly reflecting
mirror, so that the total reflected power is unchanged. 

In this paper we ask the question what happens if instead of such a normal mirror one would use a 
phase-conjugating mirror \cite{Fisher,Zeldovich}. The interplay of multiple scattering by disorder and optical 
phase conjugation is a rich problem even in the static case \cite{Yudson,Paasschens,superlat}. Here we
show that the dynamical aspects are particularly striking. Basically, the disordered waveguide is turned into a
virtual cavity with a resonance frequency $\omega_0$ set by the phase-conjugating mirror. 

We present a detailed analytical and numerical calculation for the single-mode case ($N=1$). For times $t \gg \tau_s $
we find that $a(\omega,t)$ has decayed almost completely except in a narrow frequency range $\propto \tau_s ^{-1}\exp(-L/l)$ around $\omega_0$. In this frequency range the decay is delayed up to times $t \simeq \tau_s \exp(L/l)$, after which a log-normal decay sets in. The exponentially large difference in time scales for the decay near 
$\omega_0$ and away from $\omega_0$ is a signature of localization.

\section{Formulation of the problem}

\subsection{Scattering theory}

\begin{figure}[!tb] 
\begin{center}
\includegraphics[angle=0, width=10cm]{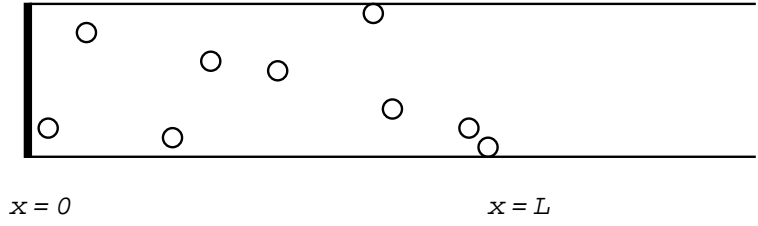}
\end{center}
\caption[x]{The geometry under investigation consists of a single-mode waveguide with a mirror at
$x=0$. It can be a normal mirror or a phase-conjugating mirror.
There are randomly positioned obstacles between $x=0$ and $x=L$.} \label {figgeometry}
\end{figure}

A scattering matrix formulation of the problem of combined elastic scattering by disorder 
and inelastic scattering by a phase-conjugating mirror was developed by Paasschens et al.~\cite{Paasschens}. We summarize the basic equations for the case of a single propagating 
mode in the geometry shown in Fig.~$\ref{figgeometry}$. A single-mode waveguide
is closed at one end ($x=0$) by either a normal mirror or by a phase-conjugating mirror. 
Elastic scattering in the waveguide is due to random disorder in the region $0 < x < L$. For simplicity we consider a single polarization, so that we can 
use a scalar wave equation. 

The phase-conjugating mirror is pumped at frequency $\omega_0$. This means that a wave incident at frequency $\omega_0+\omega$ will be retro-reflected at frequency $\omega_0 -\omega$, for $\omega \ll \omega_0$. 
For $x \gg L$ the wave amplitude at frequencies $\omega _{\pm}=\omega_0 \pm \omega$ is an incoming or outgoing plane wave,

\begin{mathletters}
%\begin{subequations}
\label{Fourierdecom}
\begin{eqnarray} 
\label{Fourdecomin}
u_{\pm}^{\mathrm{in}}(\vec{r},t)&=&{\mathrm{Re}} \hspace{0.2cm} \phi _{\pm}^{\mathrm{in}} \exp[-ik_{\pm}(x-L)-i\omega_{\pm}t]\psi_{\pm}(y,z),
\\ \label{Fourdecomout}
u_{\pm}^{\mathrm{out}}(\vec{r},t)&=&{\mathrm{Re}} \hspace{0.2cm} \phi _{\pm}^{\mathrm{out}} \exp[ik_{\pm}(x-L)-i\omega_{\pm}t]\psi_{\pm}(y,z).
\end{eqnarray}
%\end{subequations}%
\end{mathletters}%
Here $k_{\pm}=k_0 \pm \omega /c$ is the wavenumber at frequency $\omega _{\pm}$, with $k_0$ the wavenumber at $\omega _0$ and $c={\mathrm{d}}\omega /{\mathrm{d}} k$ the
group velocity.
The transverse wave profile $\psi_{\pm}(y,z)$ is normalized such that the wave carries unit 
flux. 

The reflection matrix relates the incoming and outgoing wave amplitudes, according to

\begin{equation} \label{smattotalpcm}
{\left (\begin{array} {c}
\phi_{+} \\
\phi_{-}^{*} \\
\end{array} \right)}^{\mathrm{out}}
={\left (\begin{array} {cc}
r_{++} & r_{+-} \\
r_{-+} & r_{--} \\
\end{array} \right)}
{\left (\begin{array} {c}
\phi_{+} \\
\phi_{-}^{*} \\
\end{array} \right)}^{\mathrm{in}}.
\end{equation}
The reflection coefficients are complex numbers that depend on $\omega$. They
satisfy the symmetry relations 

\begin{equation} \label{syms}
r_{--}^{*} (\omega)=r_{++} (-\omega), \hspace{0.5cm} r_{-+}^{*} (\omega)=r_{+-}(-\omega). 
\end{equation}
If there is only reflection at the mirror, and no disorder, then one has simply

\begin{equation}
{\left (\begin{array} {cc}
r_{++} & r_{+-} \\
r_{-+} & r_{--} \\
\end{array} \right)}
={\left (\begin{array} {cc}
-e^{2ik_{+}L} & 0 \\
0 & -e^{-2ik_{-}L} \\
\end{array} \right)}
\end{equation}
for a normal mirror, and

\begin{equation}
{\left (\begin{array} {cc}
r_{++} & r_{+-} \\
r_{-+} & r_{--} \\
\end{array} \right)}
={\left (\begin{array} {cc}
0 & -ie^{2iL\omega /c} \\
ie^{2iL\omega /c} & 0 \\
\end{array} \right)}
\end{equation}
for a phase-conjugating mirror operating in the regime of ideal retro-reflection. (We will
assume this regime in what follows.)

We wish to determine how the reflection coefficients are modified by the elastic scattering by the disorder. For this we need the elastic scattering matrix

\begin{equation} \label{smatopen}
S={\left (\begin{array} {cc}
r & t' \\
t & r' \\
\end{array} \right)}.
\end{equation}
The reflection coefficients $r$,$r'$ and transmission coefficients $t$,$t'$ describe reflection and transmission from the left or from the right of 
a segment of a disordered waveguide of length $L$.
The matrix $S$ is unitary and symmetric (hence $t=t'$). The basis for
$S$ is chosen such that $r=r'=0$, $t(\pm \omega)=e^{ik_{\pm}L}$ in the absence of disorder.
The relationship between the coefficients in Eqs.~($\ref{smattotalpcm}$) and ($\ref{smatopen}$) is \cite{Paasschens}

\begin{mathletters} \label{rplusplusmin}
%\begin{subequations} \label{rplusplusmin}
\begin{eqnarray} \label {eq:rplusplus}
r_{++}(\omega)&=&r'(\omega)+t(\omega)[1-r^{*}(-\omega)r(\omega)]^{-1}r^{*}(-\omega)t(\omega),
\\ \label{eq:rplusmin}
r_{+-}(\omega)&=&-it(\omega)[1-r^{*}(-\omega)r(\omega)]^{-1}t^{*}(-\omega),
\end{eqnarray}
%\end{subequations}%
\end{mathletters}%
for a phase-conjugating mirror.
For a normal mirror there is no mixing of frequencies and one has simply 

\begin{mathletters} \label{rtotnm}
%\begin{subequations} \label{rtotnm}
\begin{eqnarray} \label{rtotnmplusplus}
r_{++}(\omega)&=&r'(\omega)-t(\omega)[1+r(\omega)]^{-1}t(\omega),
\\
r_{+-}(\omega)&=&0.
\end{eqnarray}
%\end{subequations}%
\end{mathletters}%
In each case the matrix of reflection coefficients is unitary, so

\begin{equation} \label{uniscatmat}
|r_{++}(\omega)|^2+|r_{+-}(\omega)|^2=1.
\end{equation}

\subsection{Power spectrum}

We assume that a pulse $\propto \delta (t)$ is incident at $x=L$ [corresponding to $\phi_{\pm}^{\mathrm{in}}=1$ for all $\omega$ in Eq.~($\ref{Fourierdecom}$)]. The reflected wave at $x=L$ has amplitude
\begin{equation}
u_{\mathrm{out}}(t)  =  {\mathrm{Re}}\hspace{0.2cm}    e^{-i\omega_0 t} \int_{0}^{\infty}
\frac{{\mathrm{d}}\omega}{2\pi} \hspace{1ex} \bigg( \Big[ r_{++}(\omega)+r_{+-}(\omega) \Big] e^{-i\omega t} + \Big[r_{--}^{*}(\omega)+r_{-+}^{*} (\omega)\Big]e^{i\omega t}\bigg).
\end{equation}
(We have suppressed the transverse coordinates $y, z$ for simplicity of notation.) Using 
the symmetry relations ($\ref{syms}$), we can rewrite this as

\begin{equation}
u_{\mathrm{out}}(t)={\mathrm{Re}}\hspace{0.2cm}  e^{-i\omega_0 t} \int_{-\infty}^{\infty} \frac{{\mathrm{d}}\omega}{2\pi} \hspace{1ex} \Big[r_{++}(\omega)+r_{+-} (\omega)\Big]e^{-i\omega t}.
\end{equation}
The time correlator of the reflected wave becomes 

\begin{eqnarray}
u_{\mathrm{out}}(t)u_{\mathrm{out}}(t+t') &=&\case{1}{2} \hspace{0.1cm} {\mathrm{Re}} \hspace{0.2cm} e^{i\omega_0 t'} \int_{-\infty}^{\infty} \frac{{\mathrm{d}}\omega}{2\pi} \int_{-\infty} ^{\infty} \frac{{\mathrm{d}} \omega '}{2\pi} \hspace{1ex} e^{i(\omega ' - \omega) t } e^{i\omega ' t'}
\nonumber \\
&& \mbox{} \hspace{1cm} \times \hspace{1ex} \Big[ r_{++}(\omega)+r_{+-}(\omega)\Big] \Big[r_{++}^{*}(\omega ')+r_{+-}^{*}(\omega ')\Big],
\end{eqnarray}
plus terms that oscillate on a timescale $1/\omega_0$. We make the rotating wave approximation and neglect these rapidly oscillating terms. The power spectrum $a$ of the reflected wave
is obtained by a Fourier transform, 

\begin{eqnarray} \label{defpowerspec}
a(\omega,t)& = & \int_{-\infty}^{\infty} {\mathrm{d}}t' \hspace{1ex} \cos\big[(\omega_0+\omega)t'\big] u_{\mathrm{out}}(t)u_{\mathrm{out}}(t+t')
\nonumber\\
&=& {\mathrm{Re}} \int_{-\infty}^{\infty} \frac{{\mathrm{d}}\delta \omega}{2\pi} \hspace{1ex} e^{-i\delta \omega t} a(\omega,\delta \omega),
\end{eqnarray}
where we have introduced the correlator in the frequency domain

\begin{equation} \label{dwpowerspec}
a(\omega,\delta \omega)=  \case{1}{4}  \Big[ r_{++}(\omega +\delta \omega)+r_{+-}(\omega +\delta \omega)\Big] \Big[ r_{++}^{*}(\omega)+r_{+-}^{*}(\omega)\Big]. 
\end{equation}

Integration of the power spectrum over time yields, using also Eq.~($\ref{uniscatmat}$), 

\begin{equation}
\int_{-\infty}^{\infty} {\mathrm{d}}t \hspace{1ex} a(\omega,t)={\mathrm{Re}} \hspace{1ex} a(\omega,\delta \omega =0)=\case{1}{4} + \case{1}{2} \hspace{0.5ex} {\mathrm{Re}} \hspace{1ex}r_{+-}(\omega)r_{++}^{*}(\omega).
\end{equation}
For a normal mirror $r_{+-}(\omega)=0$ and $a(\omega,\delta \omega =0)=\case{1}{4}$, expressing 
flux conservation. For the phase-conjugating mirror there is
inelastic scattering, which mixes the frequency components $\omega$ and $-\omega$. 
The constraint of flux conservation then takes the form

\begin{equation} \label{newconstraint}
a(\omega,\delta \omega =0)+a(-\omega,\delta \omega =0)=\case{1}{2}.
\end{equation}
This follows from the symmetry relations ($\ref{syms}$) and the unitarity of the reflection matrix.
Eq.~($\ref{newconstraint}$) implies that $a(\omega =0,\delta \omega =0)=\case{1}{4}$.

\section{Random scatterers}
We assume weak disorder, meaning that the mean free path $l$ is much larger than the wavelength $2\pi/k_0$. The multiple scattering by disorder localizes the wave with
localization length equal to $2l$.
We consider separately the case of a phase-conjugating mirror and of a 
normal mirror. 

\subsection{Phase-conjugating mirror}

We first concentrate on the degenerate regime of small frequency shift $\omega$, and will simplify the expressions by putting
$\omega=0$ from the start. We note that $r_{++}(0)=0$, $r_{+-}(0)=-i$, as follows
from Eq.~($\ref{rplusplusmin}$) and unitarity of the scattering matrix ($\ref{smatopen}$).  Using Eqs.~($\ref{rplusplusmin}$) and ($\ref{dwpowerspec}$), we arrive at the
power spectrum in the frequency domain

\begin{equation} \label{eq:dispcmdeg}
a(0,\delta \omega)= \frac{i}{4} \bigg( r'(\delta \omega)+\Big[1-r^{*}(-\delta \omega)r(\delta \omega) \Big]^{-1} \Big[t^{2}(\delta \omega)r^{*}(-\delta \omega)-it(\delta \omega)t^{*}(-\delta \omega)\Big] \bigg) .
\end{equation}
The scattering amplitudes have the polar decomposition $r=\sqrt{R}\exp(i\theta)$, $r'=\sqrt{R}\exp(i\theta ')$, $t=i\sqrt{1-R} \exp[\case{1}{2}i(\theta +\theta ')]$, with $R, \theta, \theta '$ real functions of frequency. The phase $\theta '$ may be 
assumed to be statistically independent of $R(\pm \delta \omega), \theta (\pm \delta \omega)$, and
uniformly distributed in $(0,2\pi)$. (This is the Wigner conjecture, proven for chaotic scattering in Ref.~\cite{Brouwer}.)
In this way only the last term in Eq.~($\ref{eq:dispcmdeg}$) survives the disorder average $\langle \cdots \rangle$,

\begin{equation} \label{getpowerspecdw}
4\Big\langle a(0,\delta \omega) \Big\rangle =\left\langle \frac{t(\delta \omega)t^{*}(-\delta \omega)}{1-r^{*}(-\delta \omega)r(\delta \omega)} \right\rangle =\sum_{m=0}^{\infty} Z_m,
\end{equation}
where we have defined $Z_m=\left\langle t(\delta \omega)t^{*}(-\delta \omega)[r^{*}(- \delta \omega)r(\delta \omega)]^m\right\rangle$.

The moments $Z_m$ satisfy the Berezinskii recursion relation \cite{Berezinskii,Fre94}

\begin{equation} \label {eq:Zmpcm}
l \frac{{\mathrm{d}}Z_m}{{\mathrm{d}}L}=m^2(Z_{m+1}+Z_{m-1}-2Z_m)+(2m+1)(Z_{m+1}-Z_m) +2i \tau_s \delta \omega (2m+1) Z_m,
\end{equation}
with $\tau_s =l/c$ the scattering time. [The mean free path $l$ accounts only for backscattering, so that the scattering time in a kinetic equation would equal $\case{1}{2} \tau_s$.] The initial condition is $Z_m(L=0)=\delta_{m,0}$. 
In App.~A we derive an analytical result for $\left\langle a(0,\delta \omega) \right\rangle$ in the small frequency range $\ln (1/\tau_s \delta \omega ) \gtrsim L/l \gg 1$. It reads 

\begin{eqnarray} \label {finalpcmdeltaom}
\Big\langle a(0,\delta \omega) \Big\rangle & = & \case{1}{2} \int_{-\infty}^{\infty} {\mathrm{d}}k \hspace{1ex} ik \hspace{0.1cm} (-2i\tau_s \delta \omega )^{ik-1/2} 2^{-3ik-1/2} \Gamma ^2 (\case{1}{2} +ik) \Gamma(\case{1}{2} -ik) 
\nonumber\\
&& \mbox{} \hspace{1.7cm} \times \Gamma ^{-1}  (1+ik) \Gamma ^{-1} (ik) \exp[-(\case{1}{4}+k^2)L/l].
\end{eqnarray}
The initial decay is determined by the contributions of the poles at $k=-\frac{1}{2}i$, $-\frac{3}{2}i$, $-\frac{5}{2}i$,

\begin{equation} \label{polespcm}
\Big\langle a(0,\delta \omega)\Big\rangle =\case{1}{4}+\case{1}{4} i\tau_s \delta \omega \exp(2L/l) -\case{1}{18} \tau_s ^2 \delta \omega ^2  \exp(6L/l)+{\mathrm{O}}(\delta \omega ^3).
\end{equation}

\begin{figure}[!tb]
\begin{center}
\includegraphics[angle=0, width=12cm]{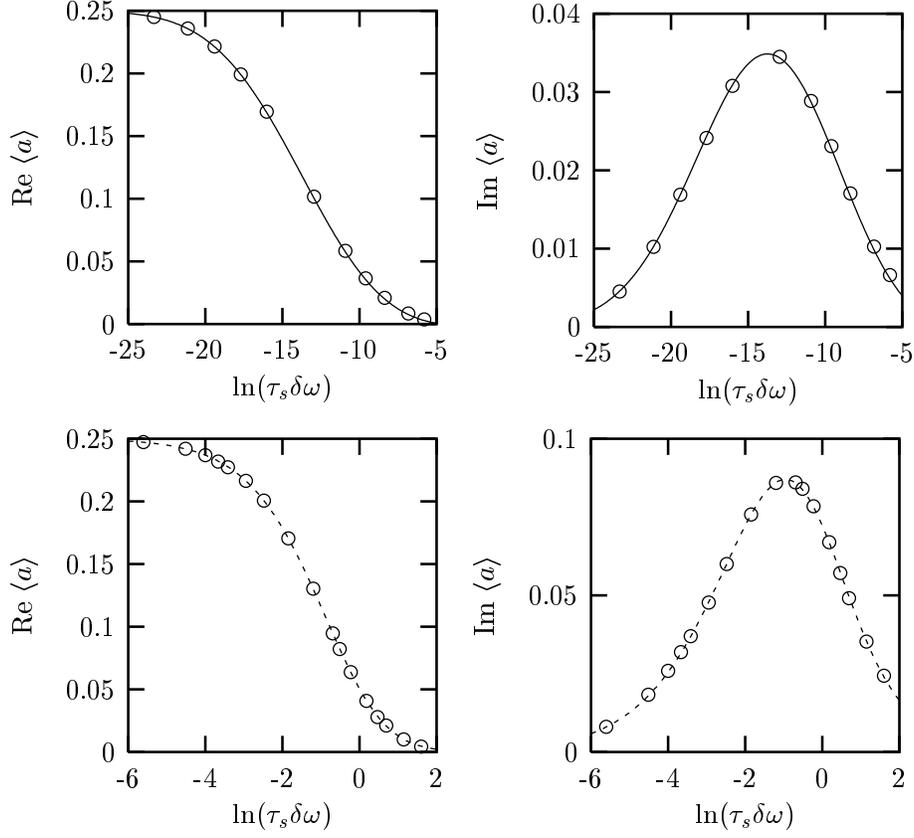}
\end{center}
\caption[x]{Average power spectrum for reflection by a disordered waveguide ($L/l=12.3$) connected to a phase-conjugating mirror [solid curves, from Eq.~($\ref{finalpcmdeltaom}$)] or a normal mirror [dashed curves, from Eq.~($\ref{infresdw}$)]. The data points follow from a numerical simulation. There is no adjustable parameter in the comparison. Notice the much faster frequency dependence for the phase-conjugating mirror (top panels), compared to the normal mirror (bottom panels).} \label {figpcmnm}
\end{figure}

The result ($\ref{finalpcmdeltaom}$) is plotted in Fig.~$\ref{figpcmnm}$ for $L/l=12.3$. We compare with the data
from a numerical solution of the wave equation on a two-dimensional lattice, using the
method of recursive Green functions \cite{Bar91}. (The method of simulation is the 
same as in Ref.~\cite{Paasschens}, and we refer to that paper for a more
detailed description.) The agreement with the analytical 
curves is quite good, without any adjustable parameter. The $\delta \omega $-dependence of $\langle a(0,\delta \omega) \rangle$ for large $L/l$ occurs on an
exponentially small scale, within the range of validity of Eq.~($\ref{finalpcmdeltaom}$).

A Fourier transform of Eq.~($\ref{finalpcmdeltaom}$) yields the average power spectrum in the time domain for $\ln(t/\tau_s) \gg  L/l \gg 1$, with the result 

\begin{equation} \label{finaldispcmtime}
\Big\langle a(0,t)\Big\rangle =\case{1}{8} \pi^{3/2} (L/l)^{-3/2} \exp(-L/4l) \tau_s^{-1/2} t^{-1/2} \ln (4t/\tau_s) \exp\Big[-(l/4L)\ln ^2 (4t/\tau_s)\Big].
\end{equation}
The leading logarithmic asymptote of the decay is log-normal $\propto \exp[-(l/ 4L) \ln ^2 t]$, characteristic of anomalously localized states \cite{dynAltshuler,AKL,Muzy,Bolton,Mirlin}.

These results are calculated for $\omega =0$, and remain valid as long as $\omega \ll \tau_s ^{-1} \exp(-L/l)$.
This is the degenerate regime.
For larger frequency mismatch $\omega$ one enters the non-degenerate regime.
The power spectrum in that regime is the same as for a normal mirror, calculated
in the next subsection. 

\subsection{Normal mirror}

For comparison we discuss the known results for a disordered waveguide connected to a normal 
mirror instead of a phase-conjugating mirror. Since $r_{+-}=0$, one has from Eq.~($\ref{dwpowerspec}$)

\begin{equation}
4 \Big\langle a(\omega,\delta \omega) \Big\rangle = \Big\langle r_{++}(\omega +\delta \omega)r_{++}^{*}(\omega)\Big\rangle \equiv R_1.
\end{equation}
The quantities $R_m=\left\langle [r_{++}(\omega+\delta \omega)r_{++}^{*}(\omega)]^m\right\rangle$ satisfy the Berezinskii recursion relation \cite{Berezinskii,Fre94}

\begin{equation} \label{recurnm}
l \frac{{\mathrm{d}} R_m}{{\mathrm{d}} L}=m^2(R_{m+1}+R_{m-1}-2R_m)+2i\tau_s \delta \omega m R_m.
\end{equation}
The initial condition is $R_m(L=0)=1$ for all $m$. The solution for $\ln(1/\tau_s \delta \omega ) \gtrsim L/l$ is known \cite{Altshuler}, and gives the 
average power spectrum 

\begin{eqnarray} \label{finalnmdw}
&& \Big\langle a(\omega,\delta \omega)\Big\rangle =\case{1}{2} \sqrt{-2i\tau_s \delta \omega } \bigg ( {\mathrm{K}}_1 \bigg[ 2\sqrt{-2i\tau_s \delta \omega } \bigg]
\nonumber\\
&& \;\; \mbox{} + \frac{1}{\pi} \int_{-\infty}^{\infty} {\mathrm{d}}k \hspace{1ex} k \sinh (\pi k) (\case{1}{4}+k^2)^{-1} {\mathrm{K}}_{2ik}\bigg[ 2\sqrt{-2i\tau_s \delta \omega }\bigg] \exp \Big[-(\case{1}{4} +k^2)L/l \Big] \bigg),
\end{eqnarray}
with $\mathrm{K}$ a Bessel function. [The result ($\ref{finalnmdw}$) does not require $L/l \gg 1$, in contrast to Eq.~($\ref{finalpcmdeltaom}$).] The initial decay is dominated by the contributions of the poles at $k=-\frac{1}{2}i$, $-\frac{3}{2}i$, $-\frac{5}{2}i$, 

\begin{equation} \label{polesnm}
\Big\langle a(\omega,\delta \omega)\Big\rangle=\case{1}{4}+\case{1}{2}i \tau_s \delta \omega L/l-\case{1}{4} \tau_s ^2 \delta \omega ^2  \exp(2L/l)+{\mathrm{O}}(\delta \omega ^3).
\end{equation}

Comparison of Eqs.~($\ref{finalnmdw}$) and ($\ref{polesnm}$) with Eqs.~($\ref{finalpcmdeltaom}$) and ($\ref{polespcm}$) shows that the decay is much slower for a 
normal mirror than for a phase-conjugating mirror. The characteristic frequency scale is larger by a factor $\exp(2L/l)$. So Eq.~($\ref{finalnmdw}$) is not sufficient to describe the entire decay of $\langle a(\omega,\delta \omega) \rangle $, which occurs in the range $\tau_s \delta \omega \lesssim 1$.
The decay in this range is obtained by putting the left-hand-side of Eq.~($\ref{recurnm}$) equal to zero, leading to
\cite{Whi87,Gorkov}

\begin{equation} \label{infresdw}
\Big\langle a(\omega,\delta \omega)\Big\rangle =\case{1}{4}-\case{1}{2}i\tau_s \delta \omega \exp(-2i\tau_s \delta \omega) {\mathrm{Ei}}(2i\tau_s \delta \omega),
\end{equation}
where $\mathrm{Ei}$ is the exponential integral function. The range of 
validity of Eq.~($\ref{infresdw}$) is $\ln(1/\tau_s \delta \omega) \ll L/l$ and $L/l \gg 1$. The result ($\ref{infresdw}$) is plotted in 
Fig.~$\ref{figpcmnm}$, and is seen to agree well with data from the numerical simulation.

For $\ln(t/\tau_s) \ll L/l$ (and $L/l \gg 1$) one can perform the Fourier transform of Eq.~($\ref{infresdw}$)
to get the average power spectrum in the time domain \cite{Whi87},

\begin{equation}
\Big\langle a(\omega,t)\Big\rangle =\case{1}{2} \tau_s (t+2\tau_s)^{-2}, \hspace{0.5cm}t > 0.
\end{equation}
It decays quadratically $\propto t^{-2}$ for $t/\tau_s \gg 1$. For exponentially long times, $t \hspace{0.8mm} \gg \hspace{0.8mm} \tau_s \hspace{0.8mm} \exp(L/l)$, one should instead perform the 
Fourier transform of Eq.~($\ref{finalnmdw}$). One finds that 
the quadratic decay crosses over 
to a log-normal decay $\propto \exp[-(l/4L)\ln^2 t]$ \cite{dynAltshuler}, the same as for the phase-conjugating mirror.

\section{Conclusion}

We have shown that the interplay of phase-conjugation and multiple scattering by disorder leads to 
a drastic slowing down of the decay in time $t$ of the average power spectrum $\langle a(\omega,t) \rangle $ of frequency components $\omega$ of a reflected pulse. The slowing down exists in a narrow frequency
range around the characteristic frequency $\omega _0$ of the phase-conjugating mirror (degenerate regime).
If $\omega$ is outside this frequency range (non-degenerate regime), the power spectrum decays as 
rapidly as for a normal mirror. 

The slowing down can be interpreted in terms of a long-lived resonance at $\omega_0$, that is induced 
in the random medium by the phase-conjugating mirror. This resonance is known from investigations
of the static scattering properties \cite{Paasschens}. The resonance is exponentially 
narrow $\propto \tau_s ^{-1} \exp(-L/l)$ in the presence of localization (with $\tau_s$ the scattering 
time, $L$ the length of the disordered region, and $l$ the mean free path). The resonance leads to the 
exponentially large differences in time scales for the decay of the power spectrum in the degenerate 
regime and the non-degenerate regime. 

We have restricted the calculation in this paper to the case of a single propagating mode, when a 
complete analytical theory could be provided. We expect that the $N$-mode case is qualitatively
similar: An exponentially large difference in time scales $\propto \exp(L/\xi)$ for the decay in the 
degenerate and non-degenerate regimes provided the medium is localized [$L$ large compared to
the localization length $\xi=(N+1)l$]. In the diffusive regime we expect $\langle a(\omega , t)\rangle$ to decay on the time scale of the diffusion time $\tau_s (L/l)^2$. The difference with the non-degenerate regime (or a normal mirror) is then a factor $(L/l)^2$ instead of exponentially large.

In final analysis we see that phase conjugation greatly magnifies the difference 
in the dynamics with and
without localization. Indeed, if there is no phase-conjugating mirror the main difference 
is a decay $\propto t^{-3/2}$ in the diffusive regime versus $t^{-2}$ in the localized regime \cite{Titov}, but the characteristic time scale remains the same (set 
by the scattering time $\tau_s$). We therefore
suggest that phase conjugation might be a promising tool in the ongoing experimental
search for dynamical features of localization \cite{LagrevE,Chabcondmat}.

\vspace{2ex}

\appendix
\section{Power spectrum in the frequency domain}

We show how to arrive at the result ($\ref{finalpcmdeltaom}$) starting from the 
recursion relation ($\ref{eq:Zmpcm}$). We assume $\ln(1/\tau_s \delta \omega) \gtrsim L/l \gg 1$.
It is convenient to work with the Laplace transform
\begin{equation}
Z_m (\lambda)=\int_{0}^{\infty} \frac{{\mathrm{d}}L}{l} \hspace{1ex} \exp(-\lambda L/l) Z_m(L)
\end{equation}
of the moments $Z_m$.
The recursion relation ($\ref{eq:Zmpcm}$) transforms into
\begin{eqnarray} \label{eq:laplaceset}
\lambda Z_m(\lambda)-\delta _{m,0} &=& m^2\Big[ Z_{m+1}(\lambda)+Z_{m-1}(\lambda)-2Z_m(\lambda)\Big] +(2m+1) \Big[ Z_{m+1}(\lambda)-Z_m(\lambda)\Big] 
\nonumber\\
&& \mbox{} -\beta (2m+1) Z_m (\lambda),
\end{eqnarray}
with $\beta =-2i\tau_s \delta \omega$. 

For small $|\beta |$ and large $m$ this equation can be written as a differential equation,

\begin{equation} \label{dvm}
m^2 \frac{\partial^2 Z(m,\lambda)}{\partial m^2}+2m\frac{\partial Z(m,\lambda)}{\partial m}-(\lambda +2\beta  m)Z(m,\lambda)=0,
\end{equation}
where $m$ is now considered to be a continuous variable.
The solution of Eq.~($\ref{dvm}$) is
\begin{equation} \label{resdvpcm}
Z(m,\lambda)=C(\lambda,\beta) (\beta m)^{-1/2} {\mathrm{K}}_{\sqrt{1+4\lambda}}\left(2\sqrt{2\beta m}\right).
\end{equation}
The factor $C(\lambda,\beta)$ is determined by matching to the solution of Eq.~($\ref{eq:laplaceset}$) for $\beta m \to 0$, $m \to \infty$, that has been calculated in Ref.~\cite{Melnikov}. The result is

\begin{eqnarray} \label{anpcmzm}
C(\lambda,\beta)&=&4 \pi  \beta ^{1/2} \Gamma \Big(\case{1}{2} +\case{1}{2} \sqrt{1+4\lambda}\Big) \Gamma ^{-1} \Big(1+\case{1}{2} \sqrt{1+4\lambda}\Big) \Gamma ^{-1} \Big(\case{1}{2} \sqrt{1+4\lambda}\Big) 
\nonumber\\
&& \mbox{} \times \exp \Big[ \case{1}{2} \sqrt{1+4\lambda}\ln(\beta/8)\Big].
\end{eqnarray}

To obtain the power spectrum ($\ref{getpowerspecdw}$) we replace the sum over $m$ by
an integration, with the result

\begin{eqnarray}
\sum_{m=0}^{\infty} Z_m (\lambda)& =& 2^{1/2} \pi  \beta ^{-1/2} \Gamma ^2 \Big(\case{1}{2}+\case{1}{2}\sqrt{1+4\lambda }\Big) \Gamma\Big(\case{1}{2} -\case{1}{2}\sqrt{1+4\lambda }\Big) \Gamma ^{-1} \Big(1+\case{1}{2}\sqrt{1+4\lambda}\Big)
\nonumber\\
&& \mbox{} \times \Gamma ^{-1} \Big(\case{1}{2}\sqrt{1+4\lambda}\Big) \exp \Big[\case{1}{2}\sqrt{1+4\lambda}\ln(\beta/8)\Big].
\end{eqnarray}
There are poles at $\lambda =n(n+1)$, $n=0,1,2,\ldots$ and a branch cut starting at $\lambda=-1/4$.
When doing the inverse Laplace transform we put the integration path in between
the poles and the branch cut. The final result is given by Eq.~($\ref{finalpcmdeltaom}$). 
The reason that we need the condition $L/l \gg 1$ is that Eqs.~($\ref{resdvpcm}$) and ($\ref{anpcmzm}$) are only correct for $m \gg 1$. The first terms in the sum $\sum_{m=0}^{\infty}Z_m$ are
important for $L/l \lesssim 1$, but can be neglected for $L/l \gg 1$.

\end{document}